\documentclass[aip,amsmath,amssymb,reprint,letterpaper]{revtex4-1}

\usepackage{graphicx}
\usepackage{color}
\usepackage{natbib}

\begin{document}

\title{Examining Convergence of Cluster Perturbation Theory and Cluster Coupled
Cluster for J1-J2 Heisenberg Spin Lattices}

\author{Jeffrey R. Keyes}
% \email{jk220@rice.edu}
\affiliation{Department of Chemistry, Wesleyan University, Middletown,
  CT 06459}

\author{Carlos A. Jim\'enez-Hoyos}
\affiliation{Department of Chemistry, Wesleyan University, Middletown,
  CT 06459}

% \date{\today}

\begin{abstract}
  In previous work, it was shown that using the cluster mean field(cMF)
  wavefunction as the zeroth order wavefunction for perturbation theory and
  coupled cluster produces a reasonable approximation for the $J_1-J_2$
  Heisenberg model in the thermodynamic limit(TDL). However, this work was
  limited to cPT4 and cCCSD. The central claim of cMF is that it takes strongly
  correlated systems and makes them weakly correlated so that PT weakly
  correlated methods can be used. To confirm that this claim is true, it is
  important to show that the PT and CC series converge. In this paper, we
  extend the previous calculations to cPT7 and cCCSDTQ5 where we show that the
  energy is converging. We also discuss how to perform cluster based
  calculations efficiently in the thermodynamic limit, which is crucial for
  these more expensive calculations.
\end{abstract}

\maketitle

\section{Introduction}
Systems with strongly correlated electrons show up quite frequently throughout
quantum chemistry and condensed matter physics such as in conjugated molecules
or transition metal complexes. Accurately calculating the electronic structure
of these systems proves to be challenging since weak correlation methods like
perturbation theory(PT) and coupled cluster(CC) do not work.  Instead,
multiconfigurational methods like complete active space self consistent
field(CAS-SCF)\cite{zhao_casscfcaspt2_2009, sayfutyarova_constructing_2019,
helmich-paris_benchmarks_2019} are used where a subspace of the Hilbert space
containing strongly correlated electronic configurations are included, then the
\\Hamiltonian is diagonalized in this subspace. A major issue with this method is
that the dimension of the active space can become unreasonably large for many
systems making the calculation intractable. For example, this happens in
molecules with large conjugated systems like porphyrins.

A recently proposed method to address this issue is to use a cluster
mean-field(cMF) wavefunction as the zeroth order wave function for PT and
CC.\cite{jimenez-hoyos_cluster-based_2015,
  papastathopoulos-katsaros_coupled_2022, papa_2023, papa_2024, mayhall_2021,
mayhall_2022, mayhall_2023, mayhall_2024, mayhall_2025} The idea is to
partition the Hilbert space into clusters each containing a subset of the
single particle states of the system. Each of these clusters can be treated
using a multiconfigurational approach where the full wavefunction is taken as a
antisymmetrized product of the cluster wavefunctions so that the clusters
interact at a mean field level. Ultimately, this method is intended to turn
strongly correlated problems into weakly correlated ones so that PT and CC can
be used to correct the interaction between clusters.  If these clusters are
chosen wisely, much of the correlation will be included in the cMF
wavefunction, then a weak correlation technique can be used to improve the
interaction of the clusters. In this paper, we examine the case where each
cluster is treated exactly with a full configuration interaction(FCI)
wavefunction.

This method was recently applied to a Heisenberg spin lattice with fourth order
perturbation theory and CCSD where clusters where chosen to group nearby
sites.\cite{papastathopoulos-katsaros_coupled_2022} These calculations were
successfully carried out for $J_2/J_1$ between 0 and 1 which converged in all
regions including the difficult paramagnetic region, where the system becomes
spin frustrated. The success of both PT4 and CCSD on the very strongly
correlated spin system indicates that the cMF wavefunction incorporated enough
correlation that the clusters are weakly correlated. Despite this success, the
question of whether the correlation between clusters was made sufficiently small
to allow the perturbation and coupled cluster series to converge remains.

The Heisenberg system has important applications in chemical systems. In
particular, the theoretical modeling of many metal clusters and molecular
magnets is based on effective spin interactions.\cite{molecularmagnets} While
many approaches have been put forward in the literature to describe the ground
state (as well as excited states) of spin lattices, our focus with this work is
to keep assessing the quality of the ansatz with an eye to an eventual
application in molecular systems.

In this paper, we extend the previous work by obtaining perturbation energies up
to PT7 and CC energies up to CCSDTQ5. These data points give us a better
indication of the convergence behavior of these series and allow us to
extrapolate to gather accurate energies for the spin system. Further, the fact
that we can converge our wavefunction in paramagnetic region means that we can
more directly determine the positions of the phase transitions from the n\'{e}el
region to the paramagnetic region and from the paramagnetic region to the
collinear region. One important difference between the previous work and this
one is that we perform all calculations directly in the thermodynamic
limit(TDL), whereas the previous work was done by setting up periodic square
lattices. This change is important when calculating high order PT energies
because the higher PT and CC terms would require larger periodic squares, which
causes the calculation to become intractable. We will discuss how to perform the
calculation in the TDL efficiently.

The remainder of this paper is organized as follows, in Section
\ref{sec:Theory}, give some background of the heisenberg system and present some
details of the cMF ansatz as well as some details of how the PT calculations wer
carried out. in Section \ref{sec:Results}, we examine our results. Finally, in
Section \ref{sec:Conclusion}, we provide some concluding thoughts and discuss
future directions for the project.

\section{Theory}\label{sec:Theory}
\subsection{Heisenberg Model}
\begin{figure}[ht]
  \begin{center}
    \includegraphics*[scale=0.5]{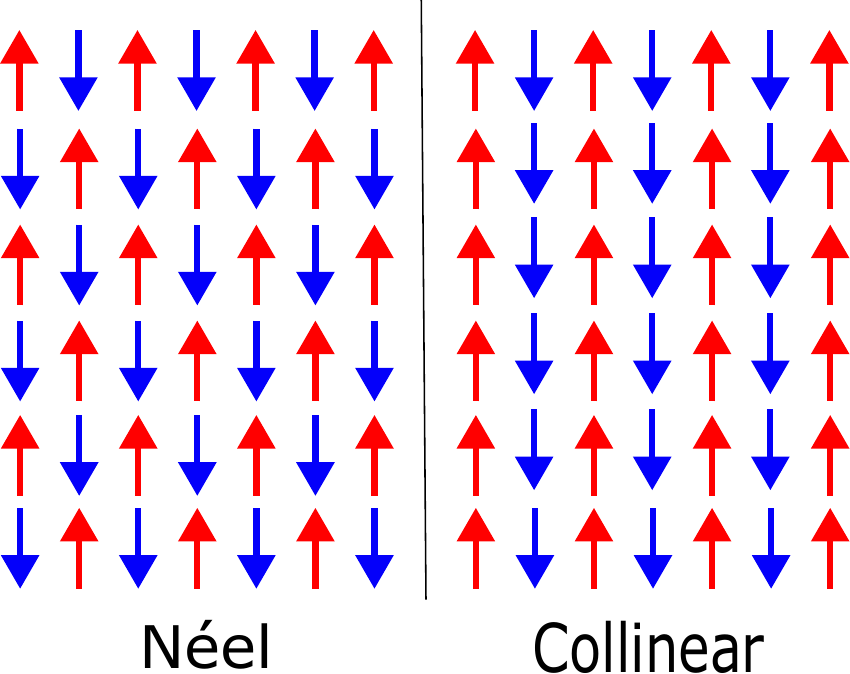}
  \end{center}
  \caption{Dominant spin configuration for $J_2/J_1 \lesssim 0.4$(left) 
  and $J_2/J_1 \gtrsim 0.6$(right)}
  \label{fig:spin}
\end{figure}

In this paper we perform calculations using the $J_1$-$J_2$ Heisenberg
model on a square grid in the thermodynamic limit (TDL). This grid
contains a spin $1/2$ particle at each point which interact with the
Hamiltonian:
\begin{equation}
  H = J_1\sum_{\langle ij\rangle}\vec{S}_i\cdot\vec{S}_j + J_2\sum_{\langle\langle ij\rangle\rangle}\vec{S}_i\cdot\vec{S}_j
\end{equation}
where $\vec{S}_i$ is the spin operator on site $i$ and $J_1$ and $J_2$
are the coupling constants for nearest-neighbor and
second-nearest-neighbor interactions, respectively. We work in the
anti-ferromagnetic regime with $J_1,J_2 > 0$ and we further focus on
$J_1 \geq J_2$. The nature of the ground state changes as the ratio
$J_2/J_1$ varies from 0 to 1. When $J_2/J_1 \lesssim 0.4$, the
dominant configuration is the N\'{e}el state and when $J_2/J_1 \gtrsim
0.6$, the dominant configuration is a collinear arrangement of spins, as shown
in Fig. \ref{fig:spin}. In the region between these, the paramagnetic region,
there is a competition between these two configurations. There is a second
order critical point between the N\'{e}el and paramagnetic regions and a first
order transition point between the paramagnetic and collinear regions.

This system will be a good starting point for two reasons, first is that the
system is more strongly correlated than most molecules, meaning that if cluster
methods are effective for the Heisenberg Hamiltonian, then it should also be
effective for molecular systems. The second reason is that the Heisenberg
Hamiltonian is made up of bosonic spin operators and does not include any charge
transfer configurations, this makes the implementation of cluster methods
simpler since we do not need to consider commutation rules. 

\subsection{Cluster Mean Field}
The formalism of cMF has been discussed in some detail
elsewhere,\cite{jimenez-hoyos_cluster-based_2015} here we present the main
ideas and notation. The premise of cMF is to partition the basis states into
clusters containing a subset of those one particle states. In the case of the
spin systems, this amounts to partitioning the lattice sites into clusters.
Each cluster is then treated separately while keeping the interaction of
clusters at the mean field level. This can be conceptualized as a CAS-SCF
calculation with multiple active spaces. In general, the number of particles
placed in each cluster is arbitrary, but the spin system restricts the Fock
space to only use states that have a single spin on each lattice site, so the
number of particles in each cluster will be determined by the number of sites
contained in the cluster.

For the remainder of this paper, we assume that all clusters contain $N_c$
sites with no overlap; the particular arrangement of sites considered is
discussed below. We also assume translational invariance of clusters: a cluster
constitutes a unit cell that perfectly tiles the full lattice. We choose to
preserve $S_z$ as a good quantum number for the ground state of each cluster,
but we do not require that the cluster states are eigenfunctions of $S^2$ to
allow for long range magnetic ordering.

To build the wavefunction we define operators $A^\dagger_{p,c}$ $(A_{p,c})$,that
create (annihilate) the $p$-th many-particle state on cluster $c$ taken as a
linear combination of the $2^{N_c}$ spin states on cluster $c$. These states can
be ordered by ascending energy such that $A_{0,c}$ is the annihilation operator
for the ground state of cluster $c$. The cMF wavefunction is then defined as:
\begin{equation}
  |\Phi_0\rangle = \prod_{i=1}^NA_{0,i}^\dagger|-\rangle
\end{equation}
where $N$ is the number of clusters and $|-\rangle$ is a putative
vacuum.  These mean field operators are then optimized to minimize the
energy $\langle\Phi_0|H|\Phi_0\rangle$. These operators inherit their
commutation rules from the underlying particle operators, which can be
quite complicated for fermionic systems. However, the Heisenberg
Hamiltonian is constructed out of spin operators, which commute with
each other as they act on different sites. It follows that the
many-particle operators in different clusters also commute with each
other.

This formalism gives us a convenient way to build the full Hilbert by exciting
clusters out of the reference state, $|\Phi_0\rangle$. For example, we can
produce a singly excited state as:
\begin{equation}
  |\Phi_c^a\rangle = A^\dagger_{a,c}A_{0,c}|\Phi_0\rangle
\end{equation}
This can be extended to higher excitations and when all excitations up to order
$N$ are considered, the basis is equivalent to the full CI basis.

Notice that if we draw a comparison of these cluster states and the molecular
orbitals used in typical applications, there is only one occupied cluster state.
This fact will prove useful in the implementation of PT and CC, which will be
elaborated about below.

\subsection{Perturbation Theory}
We can use Rayleigh-Schr\"odinger perturbation theory (RS-PT) to
improve the quality of the cMF wavefunction.  The unperturbed
Hamiltonian for cluster $c$ takes the form:
\begin{equation}
  H_{0,c} = \sum_{i,j\in c}J_{ij} \, \vec{S}_i\cdot \vec{S}_j  + 
  \sum_{i\in c}\sum_{j\notin c} J_{ij} \, \vec{S}_i\cdot\langle\vec{S}_j\rangle
\end{equation}
with $J_{ij}$ being the coupling constant for spins $i$ and $j$. The full
unperturbed Hamiltonian is then $H_0 = \sum_c H_{0,c}$ which then defines the
perturbation $\hat{V} = H - H_0$. Since our choice of cluster states allow for
full flexibility within each cluster, the perturbation includes only terms that
couple different clusters which will recover details of the interaction between
clusters. The nth order perturbation correction to the energy can then be found
as:
\begin{equation}
  E^{(n)} = \langle\hat{V}[(E_0 - H_0)^{-1}\hat{V}]^{n-1}\rangle_L
\end{equation}
where $E_0$ is the zeroth order energy and $\langle\cdots\rangle_L$
denotes that only linked-terms are included.

Note that since the cMF optimization diagonalizes $H_0$, $(E_0-H_0)$ is
trivially inverted and due to the translational invariance, the energies for
each cluster are the same, so the energies of only one cluster needs to be
carried. It follows that the evaluation of the perturbation is most
conveniently carried out in the same basis. The perturbation in the basis
discussed above is:
\begin{equation}
  V = \sum_{\mu}^N\sum_{\nu}^{N_n}\sum_{pqrs}V_{pqrs}^{\mu\nu}A_{p,\mu}^\dagger A_{q,\nu}^\dagger A_{s,\nu}A_{r,\mu}
\end{equation}
where $N_n$ is the number of neighboring clusters that have non-zero values with
cluster $\mu$. Writing the perturbation this way allows us the advantage of using
diagrammatic techniques, which prove to make the calculation very efficient.

Here, the perturbation requires only two cluster terms, but in general, two
body interactions will have three and four cluster terms. For the Heisenberg
system, only two-body interactions are included as the Hamiltonian only couples
two spins at a time (and thus at most two clusters).

\subsection{Coupled Cluster}
As an alternative to perturbation theory, coupled cluster can be used to improve
the description of the interaction between clusters. CC is characterized by the use of 
the exponential ansatz:
\begin{equation}
  |\Psi\rangle = e^T|\Phi_0\rangle
\end{equation}
where $T$ is the cluster operator defined:
\begin{equation}
  T = \sum_{m=1}^MT_m
\end{equation}
with 
\begin{equation}
  T_m = \sum_{c\in S_m}\sum_{a_1}\sum_{a_2}
  \cdots\sum_{a_m} t_{c_1c_2\cdots c_m}^{a_1a_2\cdots a_m}
  \prod_i^mA_{a_i,c_i}^\dagger A_{0,c_i}
\end{equation}
and $S_m = \{(c_1,c_2,\cdots,c_m)|c_1\neq c_2\neq \cdots \neq c_m\}$ is the set
of all sets of clusters. The sums over $a_i$ run over all excited states of each
cluster. For example, the doubles amplitude would be written:
\begin{equation}
  T_2 = \sum_{c\in S_2}\sum_{a}\sum_{b}
  t_{c_1c_2}^{ab}
  A_{b,c_2}^\dagger A_{0,c_2}A_{a,c_1}^\dagger A_{0,c_1}
\end{equation}
with $S_2$ being the set of all pairs of clusters. In this paper, we will be
restricting ourselves to $m=5$.

The correlation energy and amplitudes can be found by solving the following equations:
\begin{equation}
  \langle\Phi_0|e^{-T}(H-E_0)e^T|\Phi_0\rangle = E_{corr}
\end{equation}
\begin{equation}
  \label{eq:CCamp}
  \langle\Phi_{c_1c_2\cdots c_m}^{a_1a_2\cdots a_m}|e^{-T}(H-E_0)e^T|\Phi_0\rangle = 0 \:\:\forall m>0
\end{equation}
If we expand the Hamiltonian in the cluster basis, similar to how we did for the
perturbation, then we can use diagrammatic techniques to make the calculation
easier.

\subsection{Tiling}
The results of any cluster-based calculation will depend on the choice of how
the states are divided into clusters. If the clusters are too small, the
interaction between clusters may not be weak enough for perturbation theory or
coupled cluster to be effective. If they are chosen to be too large, then the
calculation becomes intractable due to the increasing number of states in each
cluster. The choice of which states get put in the same cluster is also
important, since the purpose of grouping states is to account for strong
correlation in the mean field calculation, we need to ensure that strongly
correlated states are placed together. If one operates in a local representation
of single particle states, then proximity between states can be used.

\begin{figure}[ht]
  \begin{center}
    \includegraphics*[scale=0.5]{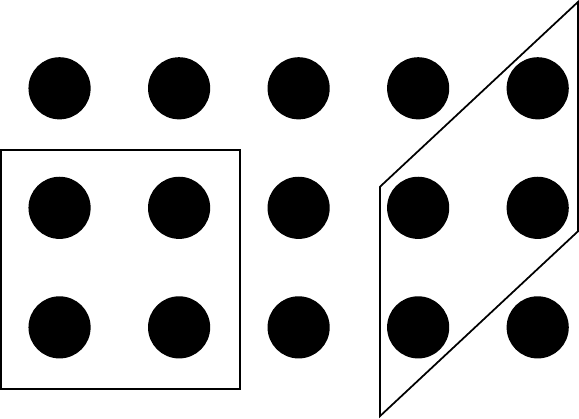}
  \end{center}
  \caption{Tile arrangements used for calculations in the N\'{e}el and paramagnetic
  regions(left) and the collinear region(right).}
  \label{fig:cells}
\end{figure}

In the case of the spin system, the effects of different cluster choices were
examined and it was found that clusters with four neighboring sites each
provided a good balance of cost and
performance.\cite{papastathopoulos-katsaros_coupled_2022} In the N\'{e}el and
paramagnetic regions, the clusters are arranged in 2x2 squares, whereas in the
collinear region, the use of tilted tiles both shown in Fig \ref{fig:cells}
provided a more robust optimization. 

With this arrangement of clusters, each cluster will interact with eight other
clusters, this limited range in the potential is what make the calculation in
the TDL possible.

Other cluster sizes were previously considered up to a 6x6 square, it was found
that increasing cluster size has a significant impact on the mean field
wavefunction, but was less important for the PT and CC energies. When attempting
to gather large order PT and CC terms, we feel it is better to go with the
smaller cluster size as using larger clusters can significantly increase the
cost of the correlated calculations. For example, in the 2x2 clusters we have 15
virtual states, so nth order amplitudes between two clusters will be carried in
sets of $15^n$, but if we use 3x3 cells then the nth order amplitudes are
carried in sets of $511^n$. So increasing the cluster size can quickly make a
calculation significantly more expensive if done without care.

Another consideration that needs to be made when deciding what clusters to use
is symmetry. Here we need to use clusters that respect the translational
invariance, otherwise we will end up with neighboring clusters that are
inequivalent. When building a symmetry into the cMF wavefunction, restricting
each cluster to contain the symmetry is an easy way to do this. As mentioned
above, we are restricting the states of each cluster to be eigenstates of the
$S_z$ operator with eigenvalue 0, but we are not doing the same for the $S^2$
operator to allow for long range magnetic ordering. This wavefunction could be
relaxed further by letting the cluster states break $S_z$ symmetry.

\subsection{Implementation}
We performed all calculations using in-house code which performs all
calculations on directly in the thermodynamic limit(TDL). For PT, this can be
accomplished by considering the perturbation written as:
\begin{equation}
  V = \sum_{\mu}^N\sum_{\nu=1}^8\sum_{pqrs}V_{pqrs}^{\mu\nu}A_{p,\mu}^\dagger A_{q,\nu}^\dagger A_{s,\nu}A_{r,\mu}
\end{equation}
where $\mu$ runs over all clusters and $\nu$ runs over the 8 neighboring
clusters of $\mu$. Note that the dependence of $V_{pqrs}^{\mu\nu}$ on $\mu$ and
$\nu$ is only on the orientation of the two clusters since all clusters are the
same by symmetry. The TDL energy is found by taking the limit
$\lim_{N\to\infty}E^{(n)}/N$, but due to the symmetry in the perturbation, if we
define:
\begin{equation}
  V' = \sum_{\nu=1}^8\sum_{pqrs}V_{pqrs}^{0\nu}A_{p,0}^\dagger A_{q,\nu}^\dagger A_{s,\nu}A_{r,0}
\end{equation}
where we have labeled some cluster 0, then the TDL energy can be found as:
\begin{equation}
  E^{(n)}_{TDL} = \langle V'[(E_0 - H_0)^{-1}V]^{n-1}\rangle_L
\end{equation}
A practical implementation of this can be realized in the following way, first use 
diagrammatic techniques to generate all PT equations. For example, one of the PT4 
terms will read:
\begin{equation}
  \sum_{\mu\nu\sigma\lambda}\sum_{abcd}\sum_{ijkl}
  \frac{V_{ijab}^{\mu\nu}V_{akic}^{\mu\sigma}V_{bljd}^{\nu\lambda}V_{cdkl}^{\sigma\lambda}}
  {\varepsilon_{ij}^{ab}\varepsilon_{jk}^{bc}\varepsilon_{kl}^{cd}}
\end{equation}
Here, $ijkl$ are the indices over occupied states, which is just the one lowest
energy state for each cluster allowing us to set $i=j=k=l=0$. The indices $abcd$
are the virtual states and run over all excited states of a cluster. Indices
$\mu\nu\sigma\lambda$ are labels for the different clusters where we have used
repeated labels where the state indices have repeated, for example indices $i$
and $a$ will be in cluster $\mu$, so that label gets repeated for the first two
operators. The term $\varepsilon_{ij}^{ab} = \varepsilon_{i} + \varepsilon_{j} -
\varepsilon_{a} - \varepsilon_{b}$ is the difference between state energies in
each cluster. Due to the translational invariance of the mean field solution,
all clusters have the same state energies so there is no need for cluster labels
in the denominator. To use this in the TDL calculation we simply need to set the 
cluster label for the first operator to 0:
\begin{equation}
  \sum_{\nu\sigma\lambda}\sum_{abcd}
  \frac{V_{00ab}^{0\nu}V_{a00c}^{0\sigma}V_{b00d}^{\nu\lambda}V_{cd00}^{\sigma\lambda}}
  {\varepsilon_{00}^{ab}\varepsilon_{00}^{bc}\varepsilon_{00}^{cd}}
\end{equation}

To limit the sums over the remaining cluster indices, we can take advantage of
the of the finite range of the perturbation. The first two operators indicate
that clusters $\nu$ and $\sigma$ must be in contact with the reference cluster.
Further, that last two operators indicate that there must exist a cluster,
$\lambda$, that is in contact with both clusters $\nu$ and $\sigma$, which
substantially restricts the summations over cluster indices. The precise
combinations of clusters can be determined with basic graph theoretical
techniques. Something worth mentioning, the first and last term in the
denominator always have the same indices as the first and last operator
respectively, so carrying their quotient as an intermediate is useful. 

For CC, the TDL limit can be reached by again defining some cluster as the
reference cluster, we then restrict the energy evaluation to only include
singles amplitudes on the reference cluster and doubles amplitudes that connect
the reference cluster to the others.

\begin{equation}
 E_{corr}^{TDL} = \frac{1}{2}\sum_{\mu}^8 V_{00ab}^{0\mu}(t_{0\mu}^{ab}+t_0^at_\mu^b)
\end{equation}

\begin{figure}[ht]
  \begin{center}
    \includegraphics*[width=0.5\textwidth]{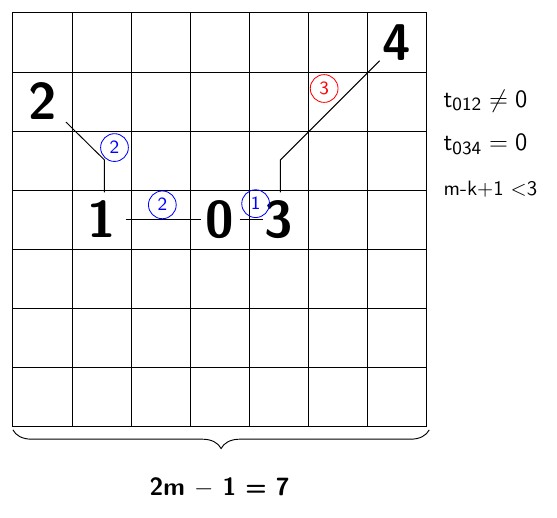}
  \end{center}
  \caption{Diagram showing an amplitude that is included and one that is not in
  the approximation for m=4 and k=3.}
  \label{fig:t_amp}
\end{figure}

Unfortunately, unlike PT, this evaluation does not reduce the calculation to be
finite since there may still be an infinite number of amplitudes that need to
be carried when solving Eq. \ref{eq:CCamp}. To remedy this, we impose the
following approximation, which is illustrated in Fig. \ref{fig:t_amp}. For CC
with excitations up to order m, amplitudes that connect the reference cluster
to any cluster outside of the square with side length of 2m-1 clusters centered
at the reference cluster is to be assumed zero. Further, the amplitude for a
kth excitation is assumed zero if the distance between the clusters in the
excitation exceeds m-k+1 clusters. For example, in Fig. \ref{fig:t_amp}, we
have a triples amplitude with excitations on the cells labeled 0, 1, and 2,
which would be included because the max distance between clusters is 2. Whereas
the triples amplitude on tiles 0, 3, and 4 would not be include because there
is a gap of 3 between the clusters used. This approximation will make the
calculation finite, while ensuring that the calculation approaches the full CC
limit correctly. The evaluation of the terms of the CC equations can be done
using similar considerations mentioned above for PT.

For both PT and CC, we used the einsum function in the numpy package\cite{numpy} for Python
to determine an efficient tensor contraction path. The contraction was then
implemented in Fortran taking advantage of Lapack\cite{lapack} subroutines whenever
possible. One advantage that PT and CC in the cluster formalism have over the
typical implementations of PT and CC is that since the occupied space for each
cluster has only one state, many of the integrals and amplitudes can be
represented as vectors and matrices, allowing for more liberal use of Lapack's
functionality.

\subsection{Magnetization}
Aside from the energy, the magnetization at each site is a physical quantity of
interest. For any finite spin lattice, the magnetization in the ground state is
zero, regardless of the $J_2/J_1$ ratio. There are still debates as to whether
there is any magnetization in the TDL.\cite{schulz_finite-size_1992,
jiang_spin_2012, wang_constructing_2013,lante_ising_2006,li_gapped_2012} The
magnetization can be evaluated by defining the modified Hamiltonian:
\begin{equation}
  H'(\lambda) = H + \lambda\sum_i B_iS_z^i
\end{equation}
where $\lambda$ is a free parameter, $S_z^i$ is the $S_z$ operator on the ith
site and $B_i$ is the strength of the interaction at each site. Obviously, $H =
H'(0)$. This can be seen as a modification of the Hamiltonian where we turn on a
small magnetic field at each lattice site where $B_i$ is chosen to preserve the
translational invariance of the state. The magnetization can then be evaluated as:
\begin{equation}
  \langle S_z^i\rangle = \frac{1}{N}\frac{\partial \langle H'\rangle}{\partial \lambda}\bigg\rvert_{\lambda=0}
\end{equation}
which we evaluate using the finite difference method.

\subsection{Data Extrapolations}
Since both PT and CC are exact in the infinite order, it can be useful to
extrapolate the finite order calculations to estimate the exact value.
Unfortunately, no scaling theory has been proven so we must rely on empirically
derived extrapolation schemes. Previous
investigations\cite{zeng_efficient_1998,kruger_quantum_2000,
li_frustrated_2013,richter_spin-12_2015,schmalfus_quantum_2006} of the
Heisenberg system with the coupled cluster method using the LSUBn approximation
used extrapolations for the TDL energy and magnetization is:
\begin{equation}
  E(n) = a + bn^{-2} + cn^{-4}
\end{equation}
\begin{equation}
  M(n) = a + bn^{-1} + cn^{-2}
\end{equation}

The lack of a proper scaling theory is typically explained away by pointing to
how extensively this extrapolation has been studied and the fact that it
extrapolates to values similar to other high accuracy models. Since our
approximation of the CC amplitudes is similar to the LSUBn approximation, we
will utilize these extrapolations for our CC results. 

While these fitting functions were suitable for the CC results, they do not
produce good fits or extrapolations for the PT energies. Instead, we find that a
good fit requires the polynomial in $n^{-2}$ to be at least order 4. For the
calculations below, we present the results of expanding the polynomial to order
5 because it was the only order that produced a reasonable extrapolation.
Expanding to other orders including to 7th order produce unphysical
extrapolations primarily in the paramagnetic and collinear regions, so the
extrapolated results for PT should be observed with caution.

\section{Results and Discussion}\label{sec:Results} 
\begin{figure}[ht]
  \begin{center}
    \includegraphics[scale=0.55]{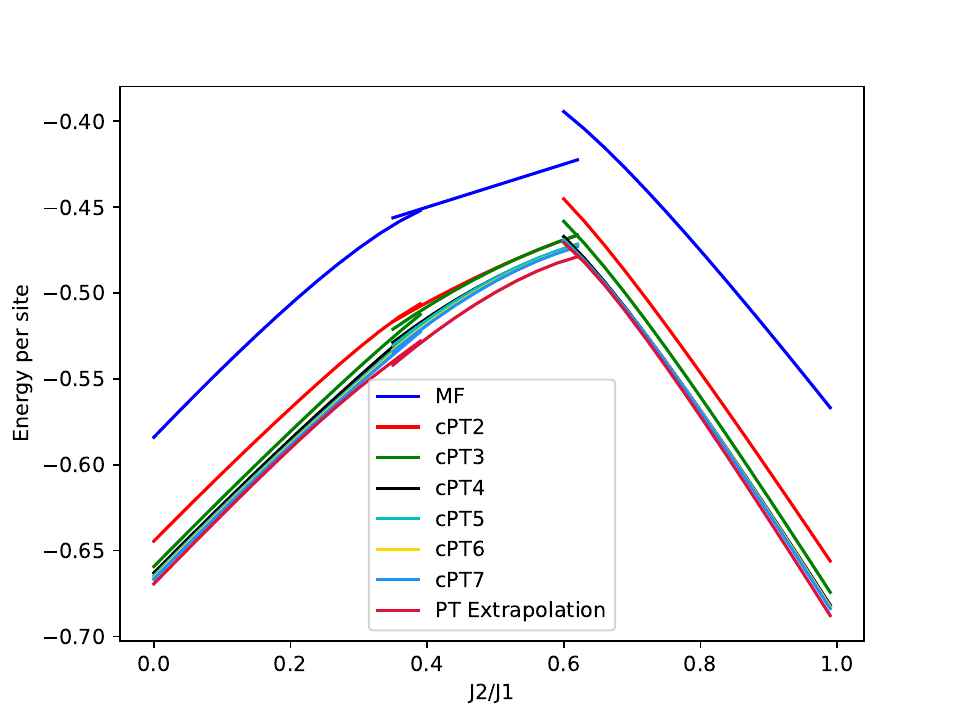}
  \end{center}
  \caption{Energy per site vs. $J_2/J_1$ for cPT calculations.}
  \label{fig:en}
\end{figure}

Our primary interest in this paper is to observe the convergence behavior of
the PT and CC series. In Fig \ref{fig:en}, we show the energy per site as a
function of $J_2/J_1$ for PT. The plot is broken up into three subplots as the
nature of the ground state changes as discussed in the introduction. The
calculation for these three sub plots differ in the arrangements of the
clusters as discussed above and the initial guess used for the mean field
calculation. The initial guess in the first plot from $J_1/J_2 \lesssim 0.4$ is
the N\'{e}el state guess whereas the plot from $J_2/J_1 \gtrsim 0.6$ uses the
collinear guess. In the intermediate region we use a paramagnetic guess where
the clusters are completely decoupled at mean-field. That is, the state in each
cluster is obtained by diagonalizing the Hamiltonian local to each individual
cluster.

From Fig. \ref{fig:en}, we find that the energy increases monotonically both the
Neel and paramagnetic regions and decreases monotonically in the collinear
region. The perturbation series seems to be converging in all regions of the
plot with the neel region roughly converging linearly with respect to $1/n^2$
and the collinear region is converging linearly after PT4. The paramagnetic
region seems to be converging the slowest where we find that odd orders of the
perturbation series contribute significantly less to the correlation energy than
the even orders, especially near the intersection of the paramagnetic and
collinear curves. This slower convergence is most likely due to the spin
frustration that occur in the paramagnetic region, increase the significance of
the correlation. We also performed an extrapolation detailed above, though we
reemphasize that the results of the extrapolation were highly dependent on the
order of the polynomial used, especially in the paramagnetic and collinear
regions. This indicates that a reliable extrapolation scheme will require either
a more detailed study of the PT energies or a proper scaling theory.

The relatively fast convergence of the correlation energy with perturbation
order is a promising indicator that the series will either converge or at worst
diverge at large order. This convergence behavior means that we chose a
perturbation that is sufficiently small, despite the strongly correlated nature
of the Heisenberg spin system. This indicates that the cluster mean field
wavefunction was capable of including enough of the correlation between
neighboring spins such that the correlation between clusters is weak.

\begin{figure}[ht]
  \begin{center}
    \includegraphics[scale=0.55]{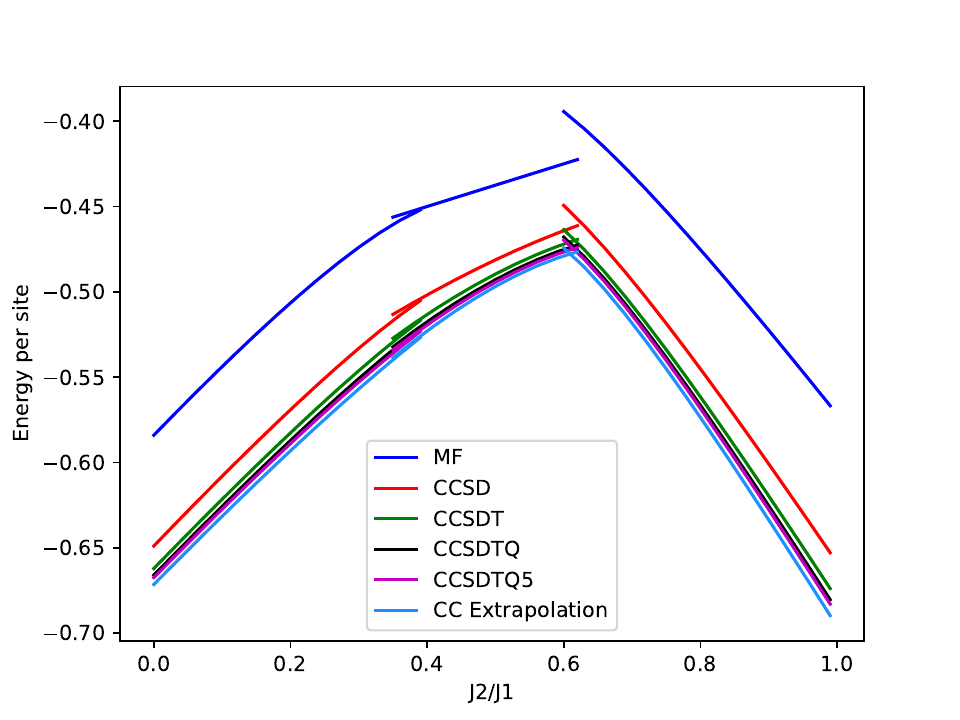}
  \end{center}
  \caption{Energy per site vs. $J_2/J_1$ for cCC calculations.}
  \label{fig:CCen}
\end{figure}

We performed the same calculations for CC up to CCSDTQ5, shown in Fig.
\ref{fig:CCen}. The features of the plot are very similar to those in the PT
plot where we note that the CC series seems to be converging faster that the PT
series. In the N\'{e}el and paramagnetic region, the CCSDTQ5 energy is about 0.1
- 1 mJ below the PT7 energies whereas in the collinear region, it is close to
the PT6 energy. We also performed an extrapolation described above, which should
be more reliable than PT due to the similarity between our CC approximation and
the LSUBn approximation. Interestingly, the CC extrapolated energies are very
similar to the PT extrapolation, the CC energy is about a mJ below the PT energy
in the N\'{e}el and collinear regions, but a mJ above the PT energy in the
paramagnetic region. 

% \begin{figure*}[ht]
%   \begin{center}
%     \includegraphics[width=\textwidth]{PT_Energy.pdf}
%   \end{center}
%   \caption{Energy per site vs. $J_2/J_1$ for cPT calculations.}
%   \label{fig:en}
% \end{figure*}
% \begin{figure*}[ht]
%   \begin{center}
%     \includegraphics[width=\textwidth]{CC_energy.pdf}
%   \end{center}
%   \caption{Energy per site vs. $J_2/J_1$ for cCC calculations.}
%   \label{fig:CCen}
% \end{figure*}


\begin{figure}[ht]
  \begin{center}
    \includegraphics[scale=0.5]{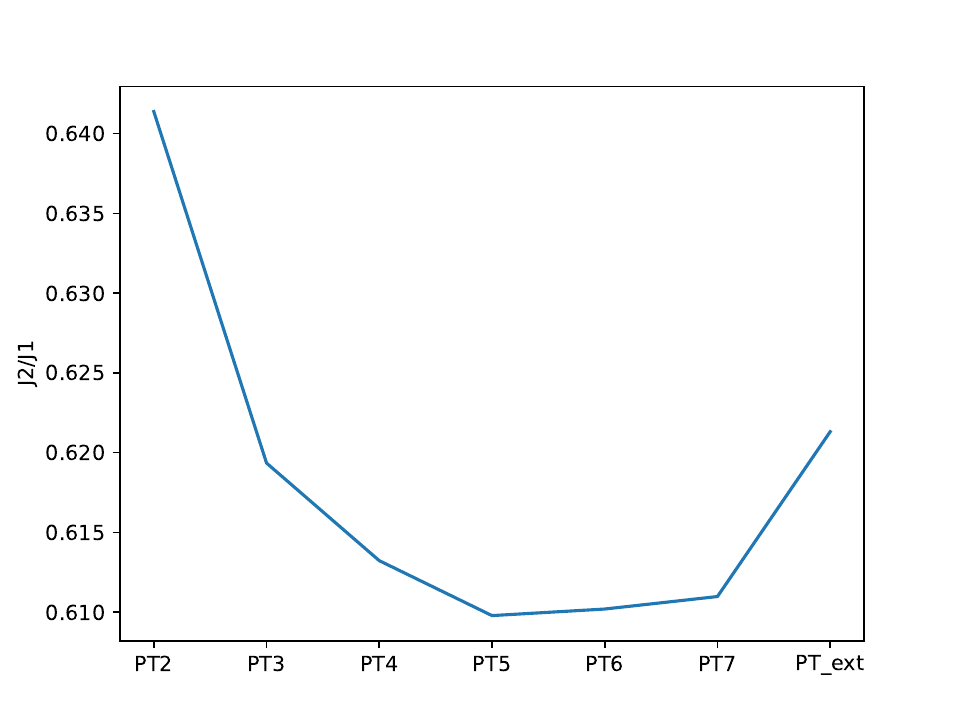}
    \includegraphics[scale=0.5]{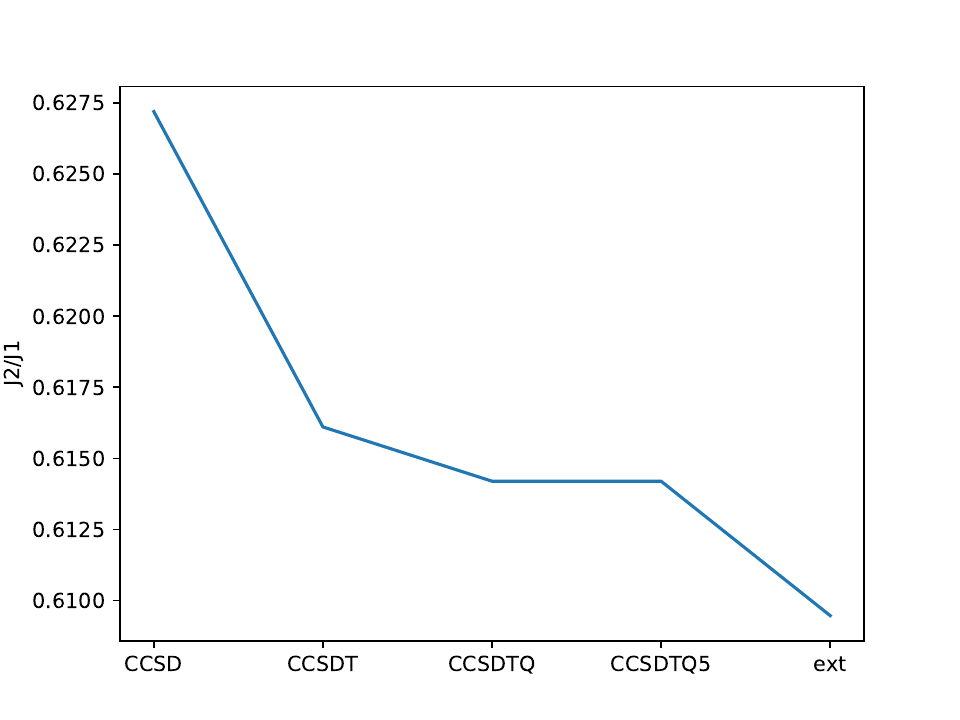}
  \end{center}
  \caption{Position of critical point for different levels of theory.}
  \label{fig:crit}
\end{figure}

Having these accurate calculations in both the paramagnetic and collinear
regions allows us to make an accurate prediction of the position of the critical
point. In Fig. \ref{fig:crit}, we show how the position of the critical point
changes for PT and CC. The PT critical point moves left in lower orders of PT,
but the moves right for PT6 and PT7. This probably due to the collinear region
having less correlation, so PT lowers the energy more at lower orders compared
with the paramagnetic region, which pushes the critical point too far left. Then
at higher orders as the paramagnetic region converges, the critical point gets
corrected. The critical point position converges monotonically for the CC
calculations which may indicate that the error in the paramagnetic and collinear
regions are more balanced in CC compared to PT. In fact, the critical point for
CCSDTQ and CCSDTQ5 are very close, so it may already be close to the converged
value. The critical point is predicted to be at 0.61098 for PT7 and 0.61420 for
CCSDTQ5. The critical points from the extrapolated energies are 0.62131 and
0.60947 for PT and CC respectively. These extrapolated values both seem to break
the patterns established by the previous orders, which is most likely caused by
differing errors in the extrapolations for the paramagnetic and collinear
regions.

\begin{figure}[ht]
  \begin{center}
    \includegraphics[scale=0.55]{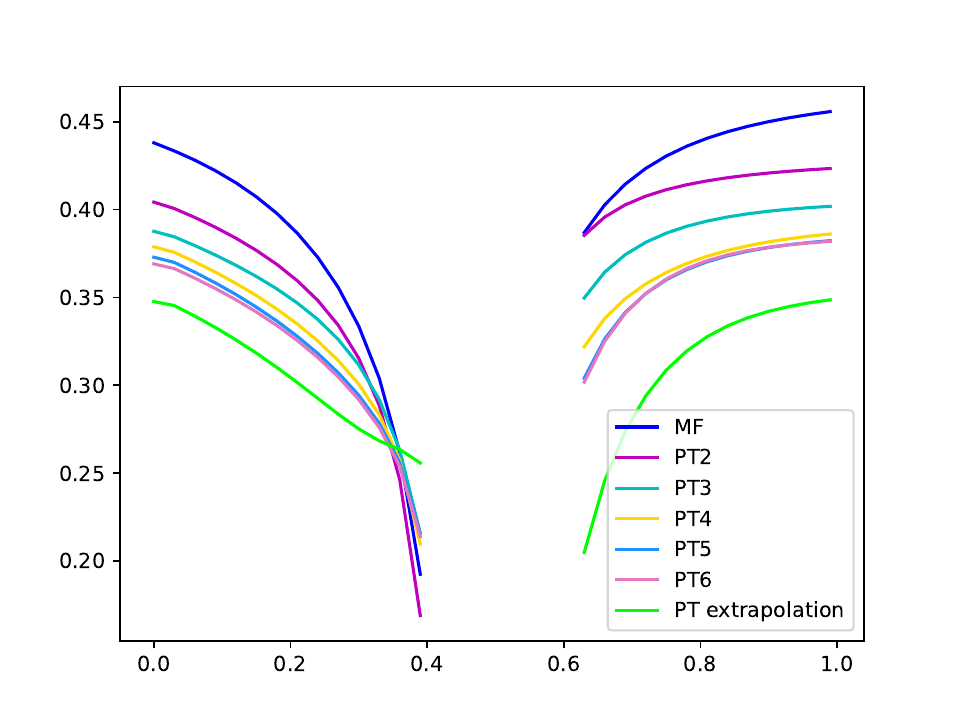}
  \end{center}
  \caption{Magnetization vs. $J_2/J_1$ for cPT calculations.}
  \label{fig:PTmag}
\end{figure}

\begin{figure}[ht]
  \begin{center}
    \includegraphics[scale=0.55]{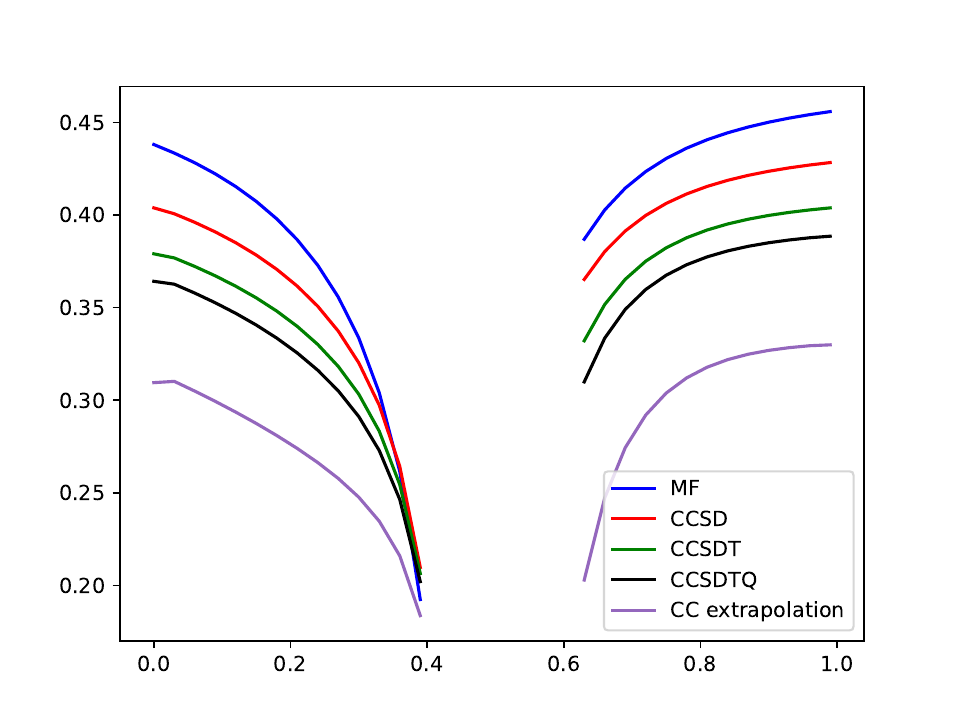}
  \end{center}
  \caption{Magnetization vs. $J_2/J_1$ for cCC calculations.}
  \label{fig:CCmag}
\end{figure}

Lastly, we present the magnetization for PT up to order 6 in Fig.
\ref{fig:PTmag} and CC up to CCSDTQ in Fig. \ref{fig:CCmag}. In both methods,
large magnetizations are predicted at $J_2/J_1 = 0$ and $J_2/J_1 = 1$ with
monotonically decreasing magnetization as the paramagnetic region is approached.
The inclusion of cluster correlation decreases the magnetization in both PT and
CC. In both methods, the numerical derivative gave unphysically low
magnetizations at $J_2/J_1 = 0.6$, which gets worse with larger order of the PT
or CC series. This is due to the fact that the magnetization of the paramagnetic
region is zero and there is a sudden drop in magnetization near the critical
point. In both PT and CC, we find that near the first critical point, the lower
order terms have a larger slope and cross the magnetization curves of the higher
orders, this may indicate that the position of the first critical point is
shifting to the right as the correlation is included. The values gathered from
both methods are very similar, the PT magnetization is about $10^{-3}$ above the
CC magnetization in the N\'eel region and $10^{-3}$ below the CC magnetization
in the collinear region. We also include extrapolations for both methods as
detailed above. These extrapolations are within 0.035 of each other except near
the first critical point where the PT extrapolation is clearly unphysical due to
the PT curves crossing.

\section{Conclusion}\label{sec:Conclusion} 
In this paper, we have performed calculations on the cluster mean field
wavefunction for the $J_1-J_2$ Heisenberg spin system with perturbation theory
up to order seven and with coupled cluster up to CCSDTQ5. We have shown that
this calculation can be carried out efficiently in the thermodynamic limit. The
results indicate that the PT and CC series are well behaved and convergent or
at worst diverging at high order, indicating that the cMF wavefunction is a
good starting point for weak correlation methods applied to strongly correlated
systems. These calculations have also given us the ability to directly find the
critical point between the paramagnetic and collinear regions of the phase
plot, giving about 0.611 for PT7 and 0.614 for CCSDTQ5. Future work for this
project include applying the same formalism to fermionic systems to deal with
strongly correlated molecular problems such as conjugated molecules, transition
metal complexes and bond breaking processes.

\section{Author Declerations}

\subsection{Conflict of Interest}
The authors have no conflicts to disclose.

\subsection{Author Contributions}
\textbf{J. Keyes}: 
Conceptualization (equal);
Data Curation (equal);
Formal Analysis (lead);
Investigation (lead);
Methodology (equal);
Software (lead);
Validation (equal);
Visualization (lead);
Writing - Original Draft (lead);
Writing - Review and Editing (equal);

\textbf{C. Jimenez-Hoyos}: 
Conceptualization (equal);
Data Curation (equal);
Funding Acquisition (lead);
Methodology (equal);
Project Administration (lead);
Resources (lead);
Supervision (lead);
Validation (equal);
Writing - Review and Editing (equal);

\section{Data Availability}
The data that supports the findings of this study are available from the
corresponding author upon reasonable request.

\nocite{*}
\bibliography{spin_paper.bib}
\bibliographystyle{plain}

\end{document}